\documentclass[12pt,aps,preprintnumbers,nofootinbib]{revtex4}

\usepackage{graphicx,amsmath,amssymb}

\renewcommand\d{\partial}

\newcommand\q{\mathbf{q}}
\renewcommand\Im{\mathop{\mathrm{Im}}}

\newcommand\<{\langle}
\renewcommand\>{\rangle}
\newcommand{\wn}{{\mathrm{w}}}
\newcommand{\qn}{{\mathrm{q}}}

\def\sumint{\hbox{$\sum$}\!\!\!\!\!\!\int}

\begin{document}

\preprint{INT-PUB-09-15}

\title{Spectral sum rules for the quark-gluon plasma}
\author{P.~Romatschke and D.~T.~Son}
\affiliation{Institute for Nuclear Theory, University of Washington, 
Seattle, Washington 98195-1550, USA}

\begin{abstract}

We derive sum rules involving the spectral density of the
stress-energy tensor in ${\cal N}=4$ super-Yang-Mills theory and pure
Yang-Mills theory.  The sum rules come from the hydrodynamic behavior
at small momenta and the conformal (in the case of ${\cal N}=4$ SYM
theory) or asymptotically free (as for the pure Yang-Mills theory)
behavior at large momenta. These sum rules may help constrain
QGP transport coefficients obtained from lattice QCD.

\end{abstract}

\maketitle

\section{Introduction}

Recently, much interest has been concentrated on the transport
properties of the strongly coupled quark-gluon plasma (QGP) created at
RHIC
\cite{Adcox:2004mh,Back:2004je,Arsene:2004fa,Adams:2005dq}.  Attempts
have been made to extract these coefficients from the lattice
\cite{Nakamura:2004sy,Aarts:2007wj,Meyer:2007ic,Meyer:2007dy,Huebner:2008as}.
These calculations rely on the reconstruction of the real-time
spectral function from Euclidean (imaginary-time) correlation
functions, which for numerical data is an ill-defined procedure unless
extra assumptions are made. In practice, the reconstruction amounts to
postulating a form of the spectral density, and then fitting the
parameters of the ansatz using lattice data.

Clearly, it would be of great help if some constraints on the spectral
density can be derived.  For nonrelativistic fluids, there exist sum
rules (for example, the $f$-sum rule) that constrain the spectral
densities~\cite{Forster}.  One may wonder if such sum rules exist in a
relativistic theory.  Some progress has been made in this direction:
for instance, Kharzeev and Tuchin \cite{Kharzeev:2007wb} (and latter
Karsch, Kharzeev, and Tuchin \cite{Karsch:2007jc}) wrote down a sum
rule for the spectral density of the trace of the stress-energy tensor.
The slope of this spectral density at zero frequency is the bulk
viscosity $\zeta$.  Although the sum rule does not fix the form of the
spectral density, with some assumptions the authors of
Refs.~\cite{Kharzeev:2007wb,Karsch:2007jc} argued that the bulk
viscosity becomes large near the QCD phase transition. (As we shall
see below, the precise form of our sum rule in the bulk channel is
slightly different from that of Kharzeev and Tuchin, but some features
of the later remain intact. We 
point out that the difference stems from a subtle non-commutativity of limits.)
Several sum rules are also argued to
hold for weakly coupled relativistic theories by
Teaney~\cite{Teaney:2006nc}.

In this paper, we derive certain sum rules for the spectral density in
hot gauge theories.  We start with the ${\cal N}=4$ supersymmetric
Yang-Mills theory, which is a prototype of the strongly coupled QGP,
and derive the following spectral sum rule:
\begin{equation}\label{shearSYM-sr}
  \frac25\, \epsilon = \frac2\pi\int\!\frac{d\omega}\omega\, 
  [\rho(\omega)-\rho_{T=0}(\omega)],
\end{equation}
where $\rho=-\Im G_R(\omega)$, and $G_R$ is the retarded propagator 
of the $T^{xy}$ component of the stress-energy tensor.  Here 
$\rho_{T=0}(\omega)$ is the spectral density at zero temperature, and 
$\epsilon$ is the finite-temperature energy density.
Besides ${\cal N}=4$ SYM theory at infinite 't Hooft coupling,
Eq.~(\ref{shearSYM-sr}) is valid for any theory whose
gravitational dual has purely Einstein gravity.  It is also valid for
${\cal N}=4$ SYM theory at any nonzero coupling.

Another sum rule relates a linear combination of second-order
hydrodynamic coefficients with the spectral function,
\begin{equation}
  \eta\,\tau_\pi - \frac12\kappa = \frac2\pi \int\limits_0^\infty\!
  \frac{d\omega}{\omega^3}\, [\rho(\omega)-\rho_{T=0}(\omega) - \eta\,\omega],
\label{shearSYM-sr2}
\end{equation}
where $\eta$ is the shear viscosity, and $\tau_\pi$ (the relaxation time)
and $\kappa$ are defined in Ref.~\cite{Baier:2007ix}.

We then move to the bulk sector of QCD and show the following sum rule
\begin{equation}\label{bulkQCD-sr}
3(\epsilon+P)(1-3c_s^2)-4(\epsilon-3 P)=
  \frac2\pi \int\!\frac{d\omega}\omega\, 
  [\rho^{\rm bulk}(\omega)-\rho^{\rm bulk}_{T=0}(\omega)],
\end{equation}
where $c_s$ is the speed of sound and 
$\rho^{\rm bulk}(\omega)$ is the spectral density for the 
trace of the energy-momentum tensor $T^\mu_\mu$.
Equation~(\ref{bulkQCD-sr}) is similar, but different, from the sum
rule suggested by Kharzeev and Tuchin, and by Karsch, Kharzeev, and
Tuchin.  We shall show that the sum rule is a consequence of the
hydrodynamic behavior of the QGP at large distances, and of asymptotic
freedom at small distances.

The structure of our paper is as follows.  In Sec.~\ref{sec:KK} we
remind the reader how spectral sum rules can be derived.  In
Sec.~\ref{sec:SYM} we derive two sum rules for the spectral function
in the shear channel.  One sum rule relates the total energy density
with the spectral density, and another sum rule relate a linear
combination of second-order hydrodynamic coefficients with the same
spectral density.  We verify both sum rules by numerically computing
the spectral integral.  We also comment on the possible form of the shear sum rule
for pure Yang-Mills theory.  In Sec.~\ref{sec:bulkQCD} we turn to pure
Yang-Mills theory and derive a sum rule for the bulk channel.

\section{Kramers-Kronig relation}
\label{sec:KK}

For definiteness, consider the retarded correlator of the $T^{xy}$
component of the energy-momentum tensor in, say, the ${\cal N}=4$ SYM
theory.  The spectral function is defined to coincide, up to a sign,
to the imaginary part of the retarded Green function of $T^{xy}$,
\begin{equation}
  \rho(\omega, \q) = - \Im G_R(\omega,\q),
\end{equation}
where we assume $\q$ to be along the $z$ direction.
Since $G_R$ is the Fourier transform of a real function (recall that
$G_R$ determines a linear response), we have 
\begin{equation}
  G_R(-\omega,\q) = G_R^*(\omega,-\q).
\end{equation}

Let us consider the function $f_\q(Z)$, defined so that
$f_\q(\omega^2)=G_R(\omega,\q)$, which has a cut from $Z=0$
to $Z=\infty$.

\begin{figure}[h]
\begin{center}
\includegraphics[width=0.4\textwidth]{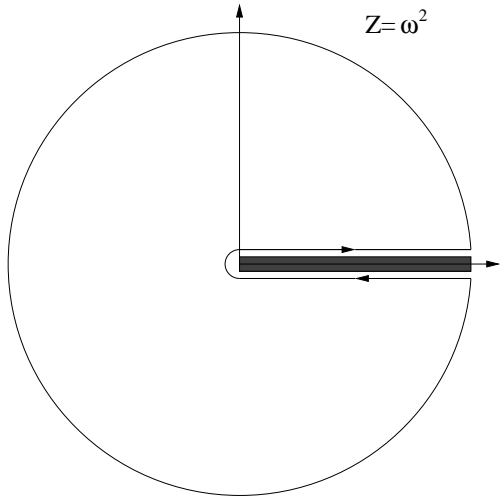}
\end{center}
\caption{Integration contour for the integral (\ref{KK1}).}
\label{fig:contour}
\end{figure}

We can write down a Kramers-Kronig relation for $f_\q(Z)$.  Pretend for
a moment that $f_\q(Z)\to0$ as $Z\to\infty$. Taking the integral of
$f_\q(Z)/(Z+\alpha^2)$ over the contour in
Fig.~\ref{fig:contour}, we find 
\begin{equation}
  \frac1{2\pi i}\oint\! dZ \frac{f_\q(Z)}{Z+\alpha^2} = f_\q(-\alpha^2),
\label{KK1}
\end{equation}
if $f_\q(Z)$ does not have singularities except for the positive real
semi-axis.  In a relativistic field theory, typically $G_R$ diverges
as $\omega\to\infty$.  For example, the $T^{xy}$ correlator
$G_R(\omega)$ grows like $\omega^4\ln\omega$ at large $\omega$ (see,
e.g., \cite{Son:2002sd}).  One can subtract this zero-temperature
piece $G_R^{T=0}(\omega)$ and we denote $\delta G_R = G_R-G_R^{T=0}$.
But, as will be shown, there remains a constant piece which
also needs to be subtracted.  Therefore, we define
\begin{equation}
  f_\q(\omega^2) = \delta G_R(\omega,\q)
  - \delta G_R^{\infty}, \qquad \delta G_R^\infty=\lim_{\omega\to+i\infty} 
  \delta G_R(\omega,\q),
\end{equation}
so that the Kramers-Kronig relation (\ref{KK1}) for this function $f_\q(Z)$
is valid. 

In a conformal, large-$N$ theory\footnote{Outside the large-$N$ limit,
there can be non-analytic terms (such as $\omega^{3/2}$) 
present in $G_R(\omega)$  \cite{Kovtun:2003vj}.},
the asymptotics of $G_R(\omega)$ at small $\omega$ and $k$ is 
known to second order from hydrodynamics \cite{Baier:2007ix},
\begin{equation}
  G_R(\omega,k) = G_R(0) - i\eta\omega 
  + \left(\eta\tau_\pi - \tfrac12\kappa\right)\omega^2 
  - \tfrac12 \kappa q^2
+{\ldots}\,.
\end{equation}
The constant $G_R(0)$ may depend on the way the correlator is defined.
When the correlator is defined through the response to metric
perturbations, $G_R(0)=P$. For this particular correlator, $f_q(Z)$
does not have singularities outside the positive real semi-axis and
the contour integral may be deformed to give
\begin{equation}
f_q(-\alpha^2)=
  \delta G_R(0) - \delta G_R^\infty +\eta\alpha 
  + \left(\eta\tau_\pi-\tfrac\kappa2\right)\alpha^2
  -\tfrac12\kappa q^2 +\ldots = -\frac2{\pi} \int\limits_0^\infty\!d\omega \,
   \frac\omega{\omega^2+\alpha^2} \delta \rho(\omega,q),
\label{specKKrel}
\end{equation}
where $\delta \rho(\omega,q)=\rho(\omega,q)-\rho_{T=0}(\omega,q)$. 

Setting $\alpha=0$ in this formula, we find
\begin{equation}
  \label{preshearSYM-sr}
  -f_q(0)=\delta G_R^\infty  - \delta G_R(0) + \tfrac12\kappa q^2 
  + {\cal O}(q^4)=
  \frac2\pi \int\limits_0^\infty\! 
  \frac{d\omega}\omega\, \delta\rho(\omega,q),
\end{equation}
which for $q=0$ will become Eq.~(\ref{shearSYM-sr}).
%
Subtracting the $\alpha$-independent part, the Kramers-Kronig relation
becomes
\begin{equation}
   \eta\alpha + (\eta\tau_\pi-\tfrac12\kappa)\alpha^2+\dots
   = \frac{2\alpha^2}\pi \int\limits_0^\infty\!\frac{d\omega}\omega\,
     \frac{\delta\rho(\omega,q)}{\omega^2+\alpha^2}\,.
\end{equation}
From this we derive another sum rule:
\begin{equation}\label{kappa-sr}
  \eta\tau_\pi - \tfrac 12\kappa + {\cal O}(q^2)= 
   \frac2\pi\int\limits_0^\infty\!\frac{d\omega}{\omega^3}\,
   [\delta\rho(\omega,q)-\eta\omega]\,,
\end{equation}
which for $q=0$ is Eq.~(\ref{shearSYM-sr2}). Note that the coefficient
$\kappa$ enters this equation, which is interesting because in the
hydrodynamic equations $\kappa$ couples only to curvature tensors
\cite{Baier:2007ix} and hence drops out for flat space.
Equation (\ref{kappa-sr}) suggests that $\kappa$ can be determined from
flat-space physics.
In fact, $\kappa$ can already be obtained from Euclidean correlators
at $\omega=0$ and small $q$.  For weakly coupled SU($N$) gauge theory,
we find a nonzero value for $\kappa$ at the lowest order of
perturbation theory (see Appendix \ref{sec:Ckappa}).  Curiously,
$\kappa$ divided by the entropy density only differs by a factor of
about two between strongly coupled ${\cal N}=4$ SYM and free SU($N$)
gauge theory.

\subsection*{A note on the definition of the correlators}

It is clear from the previous discussion that, in order to derive the
sum rule in a particular theory, one should use the same definition
for the correlation function in the UV ($\omega\to\infty$) and IR
($\omega\to0$).  In this paper, we use define the correlators through
the partition function ${\cal Z}$ in curve spacetime.  The one- and
two-point functions are given as the first and second derivatives of
$\ln {\cal Z}$ with respect to the metric.  In Euclidean signature,
one has
\begin{equation}\label{TT-grav}
  \delta \ln {\cal Z} =  \frac12\int\!dx\, \
  \<T^{\mu\nu}(x)\>\delta g_{\mu\nu}(x)
  + \frac18 \int\!dx\, dy\, 
      \<T^{\mu\nu}(x) T^{\rho\sigma}(y)\>\delta g_{\mu\nu}(x)
      \delta g_{\rho\sigma}(y) + \cdots
\end{equation}
In other words,
\begin{equation}\label{TT-gravd}
  \< T^{\mu\nu}(x) T^{\rho\sigma}(y)\> 
  = \left. 4 \frac{\delta^2 \ln{{\cal Z}}}{\delta g_{\rho\sigma}(y) \delta g_{\mu \nu}(x)} 
\right|_{g_{\alpha\beta}=\delta_{\alpha\beta}}
  = \left. 2 \frac{\delta}{\delta g_{\rho\sigma}(y)} 
  \<\sqrt{g}\, T^{\mu\nu}(x) \>\right|_{g_{\alpha\beta}=\delta_{\alpha\beta}}.
\end{equation}
An alternative definition of the correlator is through the path
integral in flat space,
\begin{equation}\label{TT-qft}
  \< T^{\mu\nu}(x) T^{\rho\sigma}(y) \>' = \frac 1{\cal Z} \int\!{\cal D}A\,
  e^{-S_E} T^{\mu\nu}(x) T^{\mu\nu}(y),
\end{equation}
where $A$ represents all fields in the theory and $S_E$ is the
Euclidean action.  The two correlators differ by a contact term\footnote{
A contact term is a term in $\<T^{\mu\nu}(x) T^{\rho\sigma}(y)\>$ that
is proportional to $\delta^4(x-y)$ or its derivatives (corresponding to
a constant or polynomial in momentum space). For theories 
which do not have derivatives of the metric in the action (such as Yang-Mills
theory), the contact terms can only be constants.}
\begin{equation}
  \<T^{\mu\nu}(x) T^{\rho\sigma}(y)\> = \< T^{\mu\nu}(x)
  T^{\rho\sigma}(y) \>' 
  - 4\left\< \frac{\delta^2 S_E}{\delta g_{\mu\nu}(x)
  \delta g_{\rho\sigma}(y)}\right\>\,.
\end{equation}
Analogously, one can define the Minkowski-space correlation functions.
The retarded Green function is found from the linear response,
\begin{equation}\label{TTR-grav}
  \< T^{\mu\nu}(x) T^{\rho\sigma}(y)\>^{\phantom{1}}_{\!R}
  = - \left. 2 \frac{\delta}{\delta g_{\rho\sigma}(y)} 
  \< \sqrt{-g}\,T^{\mu\nu}(x) \>\right|_{g_{\alpha\beta}=\eta_{\alpha\beta}}.
\end{equation}
The advantage of using the correlator defined through
Eqs.~(\ref{TT-grav}) and (\ref{TTR-grav}) is that we know this
correlator at low momenta through hydrodynamics.  Indeed, using the
hydrodynamic equations one can establish how a system responds to
external gravitational perturbations, and then use
Eq.~(\ref{TT-gravd}) to find the correlation functions (see
e.g. \cite{Son:2007vk,Gubser:2008sz}).  In addition, this is the most
natural definition that comes out of AdS/CFT correspondence.  Note
that on the lattice, so far what is normally measured
is~(\ref{TT-qft}).  However, this difference does not matter as far as
the Kramers-Kronig relation (\ref{specKKrel}) is concerned, because
subtracting $\delta G_R^\infty$ from $\delta G_R(0)$ makes the contact
term drop out from $f_q(0)$.

\section{Shear sum rules in ${\cal N}=4$ SYM and pure Yang-Mills theory}
\label{sec:SYM}

\subsection{AdS/CFT Calculation of $f_q(0)$}

For the case of large 't Hooft coupling, properties of ${\cal N}=4$ SYM
can be calculated using the AdS/CFT duality
\cite{Maldacena:1997re,Gubser:1998bc,Witten:1998qj}. In particular, it
is known how to calculate finite temperature correlators
$G_R(\omega,\q)$ in AdS/CFT \cite{Son:2007vk}. To find the
$\left\<T^{xy} T^{xy}\right\>$ correlator, we solve the equation of motion for
the $xy$ component of the metric, which is essentially the equation for a
minimally coupled scalar,
\begin{equation}
\frac{1}{\sqrt{-g}}\partial_z \left(\sqrt{-g} g^{zz} \partial_z \phi\right)
-g^{\mu \nu} k_\mu k_\nu \phi=0\,,
\label{minscal}
\end{equation}
where $\mu=0\ldots3$ indexes the usual four field theory dimensions
and $g$ is the determinant of the (five-dimensional) metric.
We denote the fifth dimension by $z$ (not to be confused
with the spatial direction in the previous section), where $z=0$
corresponds to the four dimensional boundary of AdS$_5$ space. Finite
temperature correlators in AdS/CFT can be studied by considering the
metric of a static black hole in the bulk. The location of the event
horizon $z_H$ of the black hole is related to its Hawking temperature,
$z_H=(\pi T)^{-1}$.  Being interested in $\delta G_R$ at large
imaginary $\omega\gg T$, we can restrict ourselves to the region 
of AdS space very close to the boundary.

For convenience, we shall use here the metric in Fefferman-Graham coordinates,
which has the following form \cite{Janik:2005zt} near the boundary at $z=0$,
\begin{equation}\label{FG-expanded}
  ds^2 = R^2 \frac{-dt^2+d\vec x^2+dz^2}{z^2} + \frac{R^2z^2}{4z_H^4}
  \left( 3dt^2 + d\vec x^2 \right) + O(z^4)\,,
\end{equation}
where $R$ is the scale set by the AdS radius. In these coordinates,
Eq.~(\ref{minscal}) becomes
\begin{equation}
  \phi'' - \frac3z \phi' + \left(1+ \frac{3z^4}{4 z_H^4}\right)\omega^2\phi
   - \left( 1- \frac{z^4}{4 z_H^4}\right)\q^2\phi = 0.
\label{minscal2}
\end{equation}
We will put $\q={\bf 0}$ from now on, and consider Euclidean momentum
$\omega^2=-Q^2$. In the deep Euclidean region $Q^2\to+\infty$, most of the
interesting dynamics happens near the boundary, so can
solve Eq.~(\ref{minscal2}) iteratively in inverse powers of $z_H$.  
We expand the solution as
\begin{equation}
  \phi = \phi_0 + \phi_1 + \cdots, 
\end{equation}
where $\phi_0=\frac12 (Qz)^2 K_2(Qz)$ is obtained by sending
$z_H\rightarrow \infty$ in Eq.~(\ref{minscal2}) and demanding
regularity at $z\rightarrow \infty$ (see also
\cite{Teaney:2006nc}).
The first correction $\phi_1$ satisfies
\begin{equation}
\label{minscal3}
  \phi_1'' - \frac3z \phi_1' -Q^2\phi_1 = j\equiv
  3Q^2 \frac{z^4}{4 z_H^4} \phi_0\,.
\end{equation}
The solution to this equation is formally given by a Green's function,
\begin{equation}
  \phi_1(z) = \int\!dz'\, G(z,z') j(z'),
\end{equation}
where $G(z,z')$ can be constructed from the known solutions
to the homogeneous Eq.~(\ref{minscal3}),
\begin{equation}
  f_1(z) = (Qz)^2 K_2(Qz)\,,\quad
  f_2(z) = (Qz)^2 I_2(Qz)\,,
\end{equation}
as
\begin{equation}
\label{gfeq}
  G(z,z') = -\frac1{W[f_1,f_2](z^\prime)} [f_1(z) f_2(z')\theta(z-z') +
            f_2(z) f_1(z')\theta(z'-z)]\,,
\end{equation}
where the Wronskian $W[f_1,f_2]=f_1 f_2'-f_1'f_2$ evaluates to $W[f_1,f_2](z^\prime)=Q^4 z^{\prime3}$.

Evaluating the integral (\ref{gfeq}) using 
\begin{equation}
  \int\limits_0^\infty\!dz\, z^5 K_2^2(z) = \frac{32}5\,,
\end{equation}
we then find the small $z$ asymptotics of $\phi_1$ to be given by
\begin{equation}
  \phi_1 (z) = -\frac3{10} \frac{z^4}{z_H^4}\,.
\end{equation}
Recalling that the correlation function is given by \cite{Son:2002sd}
\begin{equation}
G_R(\omega)=-\frac{N^2}{8 \pi^2} \lim_{z\rightarrow 0}
  \frac{\phi'(z) \phi(z)}{z^3}\,,
\label{corrdef1}
\end{equation}
which at $T=0$ reproduces the well known result,
\begin{equation}
  G_R^{T=0}(\omega) = - \left.\frac{N^2}{8\pi^2} \frac{\phi_0'(z)}{z^3}
  \right|_{z\to0} = \frac{N^2}{32\pi^2} Q^4\ln Q\,.
\end{equation}
The first correction due to temperature is
\begin{equation}
\lim_{\omega\rightarrow i \infty}  \delta G_R(\omega)|_{\phi_1} = 
 -\frac{N^2}{8\pi^2} \frac{\phi_1'(z)}{z^3}
  = \frac3{20\pi^2}\frac{N^2}{z_H^4}
  = \frac{3\pi^2}{20} N^2T^4  = \frac25 \epsilon \,.
\end{equation}
There are also
contributions to the correlators from the boundary terms in
the action (i.e., $\int\!d^4x \,\sqrt{-\gamma}$,
cf.~\cite{Skenderis:2002wp}) but one can check they contribute the same
amount at any $\omega$, and they are the only contribution at
$\omega=0$ (the contact terms).  Thus, in $N=4$ SYM theory one has
$f_{q=0}(0)=-\frac{2}{5}\epsilon$ and hence the sum rule
(\ref{preshearSYM-sr}) becomes
\begin{equation}
  \frac25\,\epsilon =
  \frac2\pi\int\limits_0^\infty\!\frac{d\omega}\omega\,
   [\rho(\omega)-\rho_{T=0}(\omega)]\,.
\end{equation}
We will call this sum rule the ``shear sum rule'' as the slope of
$\rho(\omega)$ at $\omega=0$ is the shear viscosity.  It is clear
from our derivation that this shear sum rule holds in any theory with
an Einstein gravitational dual.

\subsection{Rederivation of $f_q(0)$ in ${\cal N}=4$ SYM 
from OPE}
\label{sec:ope}
Within ${\cal N}=4$ SYM theory the sum rule can be derived without
relying on gravity.  We start with the operator product expansion (OPE) 
of the stress energy
tensor~\cite{Osborn:1993cr},
\begin{equation}\label{OPE}
  T_{\mu\nu}(x) T_{\rho\sigma}(0) \sim 
  C_T \frac{{\cal I}_{\mu\nu,\rho\sigma}(x)}{x^8}
  + \hat A_{\mu\nu\rho\sigma\alpha\beta}(x)T_{\alpha\beta}(0) + \cdots
\end{equation}
Here $\hat A$ contains various Lorentz structures, all scaling as
$x^{-4}$, and is given explicitly in
Ref.~\cite{Osborn:1993cr} in terms of three constants $a$, $b$, and
$c$.  In a thermal ensemble the second term averages to a constant
contribution to the correlator.  Setting $\mu=\rho=x$, $\nu=\sigma=y$,
and performing Fourier transform, we get\footnote{In
Ref.~\cite{Osborn:1993cr} the correlator is defined as the second
derivative of the partition function with respect to the {\em upper}
components of the metric, while we differentiate $\ln {\cal Z}$ with
respect to the lower components.  This difference, together with the
sign change by going from Euclidean to retarded propagator, has been
taken into account in Eq.~(\ref{GRabc}).}
\begin{equation}\label{GRabc}
  \delta G_R(\omega)|_{\omega\to i\infty} = -\frac{18(a+b)}{14a-2b-5c}P\,.
\end{equation}

In ${\cal N}=4$ SYM theory the coefficients $a$, $b$, and $c$ are
given by~\cite{Arutyunov:1999nw}
\begin{equation}
  a = - \frac{16}{9\pi^6} (N_c^2-1), \qquad
  b = - \frac{17}{9\pi^6} (N_c^2-1), \qquad
  c = - \frac{92}{9\pi^6} (N_c^2-1)\,,
\end{equation}
and hence
\begin{equation}
  \delta G_R(\omega)|_{\rm \omega\to i\infty} = \frac{11}5 P.
\end{equation}
On the other hand, the shift of $G_R$ when $\omega\to0$ can be found
from hydrodynamics, which predicts $\delta
G_R(0)=P$~\cite{Son:2007vk,Baier:2007ix}.  Therefore we find
$f_{q=0}(0)=-\frac65 P$, which corresponds to (\ref{shearSYM-sr}) since
for a conformal field theory $P=\frac13 \epsilon$.

A remark is in order. It is known that the constants $a$, $b$, and $c$
are independent of the coupling in ${\cal N}=4$ SYM theory: in fact,
their value can be found from one-loop
calculations~\cite{Arutyunov:1999nw}.  Therefore, the sum rule is
valid for any nonzero value of the coupling.

\subsection{Calculation of $f_q(0)$ in pure Yang-Mills theory}

In pure Yang-Mills theory, the UV behavior is that of a weakly coupled field
theory.  The leading terms in the OPE are the same as for free fields.
The coefficients $a$, $b$, $c$ can be found from the general
formulas~\cite{Osborn:1993cr}
\begin{subequations}
\begin{align}\label{abc-free}
  a &= \frac1{27\pi^6} n_\phi - \frac2{\pi^6} n_v,\\
  b &= - \frac4{27\pi^6} n_\phi -\frac1{2\pi^6} n_f, \\
  c &= - \frac1{27\pi^6}n_\phi - \frac1{\pi^6}n_f - \frac{8}{\pi^6}n_v,
\end{align}
\end{subequations}
where $n_s$, $n_f$, and $n_v$ are the number of real scalars, Dirac
fermions, and gauge fields in the theory.

For pure Yang-Mills theory, by repeating the calculations in section
\ref{sec:ope}, we find $f_{q=0}(0)=-2P$ (this can be checked
directly by computing the relevant Feynman diagram; see
Appendix~\ref{sec:C-pert}).  However, there is an additional subtlety
here that was not present in the previous section because the OPE of
two components of the stress-energy tensor may involve terms like
$\alpha_s F^2$, where $\alpha_s$ is the strong coupling
constant. Though formally higher order in $\alpha_s$, these terms
average to $\epsilon-3P$ which is a constant independent of the scale
$x$ in Eq.~(\ref{OPE}).  Therefore, we can tentatively write a sum rule
for pure Yang-Mills theory,
\begin{equation}
  \frac{\epsilon+P}2 + C (\epsilon -3P) = \frac2\pi\int\limits_0^\infty\!
  \frac{d\omega}\omega\, [\rho(\omega)-\rho_{T=0}(\omega)]\,.
\end{equation}
where the constant $C$ is left to be determined in by a more accurate
calculation. (However, naively applying the results of
Refs.~\cite{Novikov:1980uj} about the absence of the leading order gluon
condensate contribution in the tensor glueball channel would imply
that $C=0$).

In the large-$N_c$ limit, where second-order hydrodynamic coefficients
are well defined, the sum rule~(\ref{kappa-sr}) is valid (except that
one has to use a proper definition of $\kappa$ for nonconformal
theories~\cite{Romatschke-unpublished}).

\subsection{Numerical verification of sum rules in AdS/CFT}
\label{numads1}

In AdS/CFT, the spectral function $\rho(\omega)$ can be 
calculated numerically for arbitrary frequency/momenta from the 
solution to the mode equation (\ref{minscal}). For convenience,
we adopt a metric and coordinates such that Eq.~(\ref{minscal})
becomes \cite{Son:2002sd}
\begin{equation}
\phi^{\prime \prime} - \frac{1+u^2}{u\, f(u)} \phi^{\prime}
+\frac{\wn^2-\qn^2 f(u)}{u f^2(u)} \phi =0\,,
\label{modeq}
\end{equation}
where $u=z^2/z_H^2$, $f(u)=1-u^2$, $\wn=\omega/(2 \pi T)$, and
$\qn=q/(2\pi T)$.
Again, we set the spatial momentum $\qn=0$ in the following, although
the method described below can also handle non-vanishing momenta.

\begin{figure}[t]
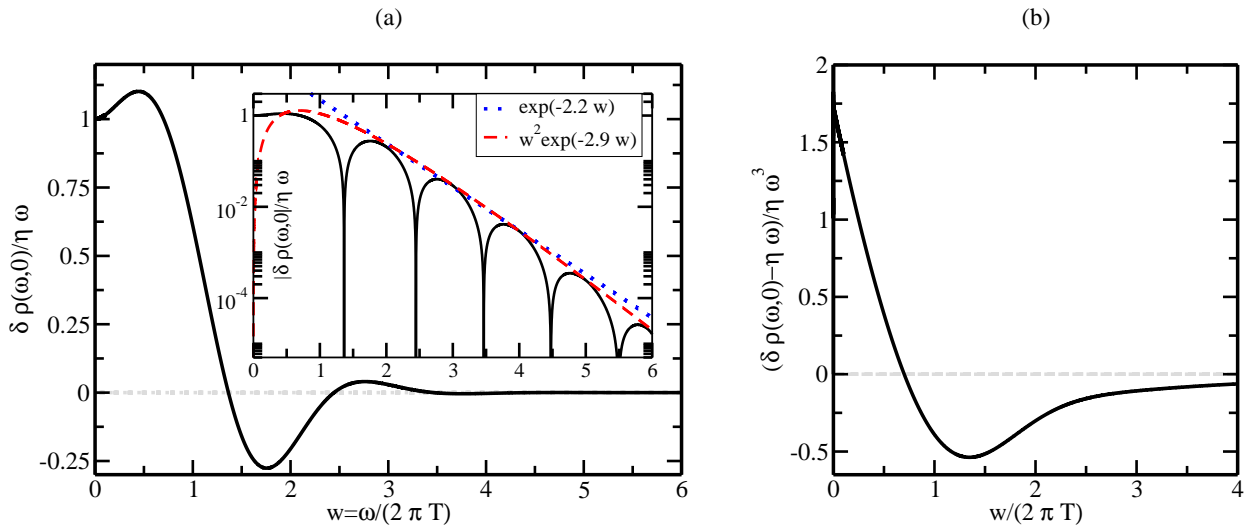

\begin{center}
\includegraphics[width=.55\linewidth]{SR1.eps}
\hfill
\includegraphics[width=.4\linewidth]{SR2.eps}
\end{center}
\caption{Numerical results for the spectral function $\delta \rho(\omega,0)$
for strongly coupled ${\cal N}=4$ SYM. The results shown correspond to the integrand
of the sum rules (\ref{shearSYM-sr}),(\ref{shearSYM-sr2}), respectively. The
inset in (a) demonstrates the near-exponential drop in the amplitude of 
$\delta \rho(\omega,0)$. Horizontal lines are visual aids to the eye.}
\label{fig:SR}
\end{figure}

We follow the algorithm by Teaney \cite{Teaney:2006nc},
which is outlined here for completeness, fixing some typos
in Ref.~\cite{Teaney:2006nc}\footnote{A version of the C++ 
code will be made available at http://hep.itp.tuwien.ac.at/\~\,paulrom}. 
Equation~(\ref{modeq}) is recast in
a system of coupled first-order equations by introducing
$\pi=\phi^{\prime}$. Discretizing derivatives as 
$\phi^{\prime}=\left(\phi(u+\delta u)-\phi(u)\right)/\delta u$,
Eq.~(\ref{modeq}) can be explicitly integrated forward from a
point close to the boundary $u=u_0$. If $\pi$ is taken to be defined
at half-integer stepsizes $\pi(u)=\pi( (n+1/2) \delta u)$ 
and $\phi$ at integer stepsizes $\phi(u)=\pi(n \delta u)$, then the
resulting algorithm is second-order accurate in $\delta u$ 
(``leapfrog algorithm''), suggesting numerical stability.
To start the algorithm, initial conditions for $\phi$ and 
$\pi$ at $u=u_0$, $u=u_0+\frac{1}{2} \delta u$, respectively,
need to be specified. For $u_0$ sufficiently close to the boundary $u=0$,
Eq.~(\ref{modeq}) may be solved analytically, yielding the pair of solutions
\begin{eqnarray}
\Phi_1(u)&=&u^2\left[1-\frac{\wn^2}{3} u
+\ldots \right],
\nonumber\\
\Phi_2(u)&=&-\frac{\wn^4}{2} \log(u) \Phi_1(u)+1+\wn^2 u-\frac{2}{9}\wn^6 u^3
+\ldots\,,
\end{eqnarray}
where the $u^2$ term in $\Phi_2(u)$ is arbitrary (it can be any
multiple of $\Phi_1(u)$) and was set to zero in accordance with the
convention by Kovtun and Starinets \cite{Kovtun:2006pf}. The analytic
result for $\Phi_{1}(u)$ (and its derivative) is used as initial
conditions for $\phi(u_0),\pi(u_0+\frac{1}{2}\delta u)$, which can
then be integrated forward to give a numerical solution $\phi_1(u_1)$
with $u_1$ close to the horizon $u=1$ (the same procedure for
$\Phi_2(u)$ gives $\phi_2(u_1)$). The physically interesting solution
for $\phi(u)$ is the one that corresponds to an incoming wave at the
horizon, $\phi(u)\sim (1-u)^{-i \wn/2}$. Solving Eq.~(\ref{modeq})
analytically close to $u=1$ one finds for the incoming wave solution
\begin{equation}
\phi^{\rm inc}(u)=(1-u)^{-i \wn/2}
 \left[1-(1{-}u)\frac{2 i \wn^3+3 \wn^2 -i \wn}{4(1+\wn^2)}
 -(1{-}u)^2 \frac{\wn(4\wn^3+7 i \wn^2-2 \wn+4 i)}{32(\wn^2+3 i \wn-2)}
 +\ldots\right].
\label{horizonsol}
\end{equation}
The real solutions $\phi_1(u)$ and $\phi_2(u)$ are linear combinations of
the incoming and outgoing wave solutions,
\begin{equation}
\begin{split}
\phi_1(u)&=A(\wn)\, \phi^{\rm inc}(u)+B(\wn)\, \bar{\phi}^{\rm inc}(u)\,,\\
\phi_2(u)&=C(\wn)\, \phi^{\rm inc}(u)+D(\wn)\, \bar{\phi}^{\rm
  inc}(u),
\end{split}
\end{equation}
where $\bar{\phi}^{\rm inc}$ denotes the complex conjugate of $\phi^{\rm inc}$.
For given $\wn$, the complex constants $A,B,C,D$ are, e.g., 
calculated from the numerical solution of
$\phi_{1,2}$ at $u=u_1$ and $u=u_1-\delta u$ and the analytic solution 
(\ref{horizonsol})
for $\phi^{\rm inc}$ close to the boundary. As a consequence, one can construct
a numerical solution to Eq.~(\ref{modeq}) with incoming wave 
boundary conditions by
\begin{equation}
\phi^{\rm inc, num}(u)=\frac{D}{AD-BC}\phi_1(u)-\frac{B}{AD - BC} \phi_2(u)\,,
\end{equation}
which can be normalized to $\phi(0)=1$ by realizing $\Phi_1(0)=0,\Phi_2(0)=1$,
so that close to the boundary
\begin{equation}
\phi^{\rm inc, norm}(u_0)=-\frac{D(\wn)}{B(\wn)}\Phi_1(u)+ \Phi_2(u)\,.
\end{equation}
In practice, we found the choices $u_0=10^{-6}$, $u_1=0.999$ to give acceptable
numerical accuracy.
Once the normalized solution to the mode equation is known, the 
retarded correlator is obtained from (\ref{corrdef1})
\begin{equation}
G_R(\omega,0)=-\frac{\pi^2 N^2 T^4}{4} \lim_{u_0\rightarrow 0} \frac{
\partial_u \phi^{\rm inc,norm}(u_0)}{u_0}\,.
\label{corrdef}
\end{equation}
In particular, using the analytical results for $\Phi_{1,2}$ one finds for the 
spectral function
\begin{equation}
\delta \rho(\omega,0)=-4 P \frac{D(\wn)}{B(\wn)}-2 \pi P \wn^4\,,
\end{equation}
where $P=\frac{\pi^2}8 N^2 T^4$ in strongly coupled ${\cal N}=4$
SYM.
The numerical result for the spectral function is shown in Fig.~\ref{fig:SR}.
As can be seen from this figure, $\delta \rho(\omega,0)$ first increases
as a function of $\omega$, reaching a maximum at around $\wn=0.45$,
then decreases strongly and oscillates around zero with an amplitude
that decays quasi-exponentially. Fig.~\ref{fig:SR}(b) shows the
spectral function where the leading hydrodynamic behavior 
$\eta\omega$ has been subtracted.
As can be seen from this figure, for small frequencies
the spectral function seems to behave as
\begin{equation}
\frac{\delta \rho(\omega,0)}{\eta \omega}=1+a_0 \wn^2+a_1 \wn^3+{\cal O}(\wn^4)\,,
\end{equation}
where numerically we determine $a_0\simeq 1.72$, $a_1\simeq -3.0$.

With the value $\frac1{8P}f_{q=0}(0)=\frac{3}{20}$ calculated above, the 
first sum rule (\ref{shearSYM-sr}) would imply the identity
\begin{equation}
\begin{split}
\frac{3}{20}=0.15&\stackrel{?}{=}
  \frac{1}{\pi}\int_0^{\infty} \frac{d \wn}{\wn} \left(
-\frac{D(\wn)}{B(\wn)}-\frac{\pi}{2} \wn^4\right)\\
  &\simeq\frac{1}{\pi}\int_0^{\wn_{\rm max}} \frac{d \wn}{\wn} \left(
-\frac{D(\wn)}{B(\wn)}-\frac{\pi}{2} \wn^4\right) =0.1500008(44),
\end{split}
\end{equation}
which we can confirm up to five digit accuracy when choosing
$\wn_{\rm max}=6$ in practice.
For the second sum rule (\ref{shearSYM-sr2}), 
$\eta \tau_\pi-\frac{1}{2} \kappa=2 P \frac{1-\log(2)}{(2 \pi T)^2}$ 
from Ref.~\cite{Baier:2007ix} implies
\begin{equation}
\begin{split}
1-\log(2)\simeq 0.306853&\stackrel{?}{=}\frac{4}{\pi}\int_0^{\infty} 
  \frac{d \wn}{{\wn}^3} \left(
-\frac{D(\wn)}{B(\wn)}-\frac{\pi}{2} \wn^4-\frac{\wn}{2} \right)\\
  &\simeq -\frac{2}{\pi \wn_{\rm max}}
  +\frac{4}{\pi}\int\limits_0^{\wn_{\rm max}}\! \frac{d \wn}{\wn^3} \left(
-\frac{D(\wn)}{B(\wn)}-\frac{\pi}{2} \wn^4-\frac{\wn}{2}\right)=0.30686(2),
\end{split}
\end{equation}
indicating that the numerical result matches with four digit accuracy.
While it is possible to improve the numerical accuracy further, we
take this agreement of at least one part in $10^{-4}$ 
between the analytical and numerical results as 
an indication that for ${\cal N}=4$ SYM, the sum rules
(\ref{shearSYM-sr}) and (\ref{shearSYM-sr2}) are correct.

\section{The bulk sum rule in QCD}
\label{sec:bulkQCD}

In this section we revisit the sum rule satisfied by the imaginary
part of the correlation function of the trace of the stress-energy
tensor $T^\mu_\mu$.  As the spectral density in this channel is
related to the bulk viscosity, this sum rule will be called the ``bulk
sum rule.''  We show that this sum rule indeed exists, but its form is
slightly different from the one given in
Refs.~\cite{Kharzeev:2007wb,Karsch:2007jc}.

In this section we shall be concerned with metric perturbations of the
following form
\begin{equation}\label{metric-scale}
  g_{\mu\nu} = \eta_{\mu\nu} e^{2\Omega},
\end{equation}
or $\delta g_{\mu\nu} = \eta_{\mu\nu} (e^{-2\Omega}-1)$, with $\Omega\ll1$.
For these perturbations, the partition function expansion 
defined in Eq.~(\ref{TT-grav}) can be explicitly given as
\begin{multline}
  \delta\ln {\cal Z}
  = \int\!dx\, \eta_{\mu\nu}(x) \<T^{\mu\nu}(x)\> 
   \left[-\Omega(x) + \Omega^2(x))\right]\\
   +\frac12 \int\!dx\,dy\, \eta_{\mu\nu}(x)\eta_{\rho\sigma}(y)
    \< T^{\mu\nu}(x) T^{\rho\sigma}(y) \> 
   \Omega(x) \Omega(y)\,.
\end{multline}
We then define the correlators of $\theta(x)$ as follows:
\begin{align}
\label{thetadef}
  \<\theta(x)\> &\equiv -\left.\frac{\delta\ln {\cal Z}}{\delta\Omega(x)}
  \right|_{\Omega=0}
  = \left. \sqrt{-g} g_{\mu \nu} \<T^{\mu\nu}(x)\> \right|_{\Omega=0}
     = \eta_{\mu\nu}(x)\<T^{\mu\nu}(x)\>\,,\\
  \<\theta(x)\theta(y)\> &\equiv \left.\frac{\delta^2\ln {\cal Z}}
     {\delta\Omega(x)\delta\Omega(y)} \right|_{\Omega=0}
  = \eta_{\mu\nu}(x)\eta_{\rho\sigma}(y)
    \< T^{\mu\nu}(x) T^{\rho\sigma}(y) \> 
    + 2\delta(x-y) \eta_{\mu\nu}(x) \<T^{\mu\nu}(x)\>\nonumber\\
  &\equiv -\left.\frac{\delta}{\delta \Omega(x)} \<\theta(y)\>
  \right|_{\Omega=0}\,,
\end{align}
where we recall that in conformal 
field theories $\<\theta(x)\>=\<\theta(x)\theta(y)\>=0$.
As a consequence of the definition (\ref{thetadef}),
the correlator $\<\theta(x)\theta(y)\>$ differs from 
$\eta_{\mu\nu}\eta_{\rho\sigma}\<T^{\mu\nu} T^{\rho\sigma}\>$ 
by a contact term, which in its turn differs
from $\<T^\mu_\mu T^\nu_\nu\>^\prime$ by a contact term.
Our subsequent calculations are simplest when using the correlator
$\<\theta(x)\theta(y)\>$ defined in this fashion.

Consider the pure Yang-Mills theory.  We need to know how to couple Yang-Mills 
to an external metric
perturbation of the form (\ref{metric-scale}).  This is done through changing the
bare coupling $g_s^2=4 \pi \alpha_s$, so that it is dependent upon the metric.
In particular, the Euclidean action of pure Yang-Mills becomes 
\begin{equation}
  S_E = \int\!dx\, \frac1{4g_s^2(\Lambda e^\Omega)} F_{\mu\nu}^2\,,
\end{equation}
where we have rescaled the gauge fields so that the field strength
tensor is given by
\begin{equation}
F_{\mu \nu}^a=\partial_\mu A_\nu^a-\partial_\nu A_\mu^a 
 + f^{abc} A^b_\mu A^c_\nu\,,
\end{equation}
and $f^{abc}$ are the $SU(N)$ structure constants.
From this we find
\begin{equation}
  \frac{\delta S_E}{\delta\Omega} 
  = \beta(g_s) \frac{\d S}{\d g_s} 
  = - \frac{\beta(g_s)}{2g_s^3} F_{\mu\nu}^2\,,
\end{equation}
where $\beta(g_s)=\Lambda \partial_\Lambda g_s$ is the beta function. 
As a consequence, one has
\begin{equation}
  \<\theta(x)\> =
  -\frac{\beta(g_s)}{2g_s^3} \< F_{\mu\nu}^2(x)\>\,,
\end{equation}
\begin{equation}\label{tt-UV}
  \<\theta(x)\theta(y)\> = \left(\frac{\beta(g_s)}{2g_s^3}\right)^2
    \<F^2(x) F^2(y)\>
  + \beta(g_s)\frac{\d}{\d g_s} \left( \frac{\beta(g_s)}{2g_s^3}\right)
   \< F^2\> \delta(x-y)\,.
\end{equation}
If we are interested in computing $G_R(\omega)$, it is most convenient to
choose $\Lambda\sim\omega$, so the two terms in Eq.~(\ref{tt-UV}) can be
evaluated perturbatively without large logarithms.
In the weak coupling regime
\begin{equation}
   \beta(g_s) = -b_0 g_s^3 
   - b_1 g_s^5 + \cdots,
\end{equation}
so the first term in Eq.~(\ref{tt-UV}) is proportional to
$g_s^4(\omega) T^4$, while the second term is proportional to
$g_s^4(\omega) (\epsilon-3P)$.  
When $\omega\to\infty$,
the correlation function vanishes because of asymptotic freedom.

Now let us compute $\<\theta\theta\>^{\phantom{1}}_{\!R}$ at small
frequencies.  For that we need to find the response of the system of
an external metric perturbation with $\Omega=\Omega(t)$ varying slowly
with $t$.  Since the perturbation is spatially homogeneous, we expect
the fluid to remain at rest ($u^0=e^{\Omega}$, $u^i=0$), but the
temperature will have time dependence: $T=T(t)$.  It is more
convenient to work with the entropy density $s$ instead of $T$.  When
metric perturbations are slow, entropy is conserved.  The solution to
the equation for entropy conservation, $\nabla_\mu(su^\mu)=0$, is
\begin{equation}
  s = e^{3\Omega} s_0\,,
\end{equation}
and therefore
\begin{equation}
  \frac{\d}{\d\Omega} = 3s \frac{\d}{\d s}\,.
\end{equation}
This result may now be directly applied to the definition of
the correlation function, and hence we find
\begin{equation}
  \< \theta \theta\>^{\phantom{1}}_{\!R}
  (\omega\to0, \q=0) = \frac{\d}{\d\Omega}\<\sqrt{-g}T^\mu_\mu\>
  =-\left(3s \frac{\d}{\d s}-4\right)(\epsilon-3P)\,.
\end{equation}

Now let us derive the spectral sum rule.  Introducing the spectral
function $\rho^{\rm bulk}(\omega)$ in the bulk channel, we find
\begin{equation}
\label{prebulk}
   \left( 3s \frac{\d}{\d s}-4\right)(\epsilon-3P)
=\frac2\pi\int\limits_0^\infty\!\frac{d\omega}\omega\,
   \delta\rho^{\rm bulk}(\omega)
\,,
\end{equation}
where $\epsilon$ and $P$ are now the thermal parts of the energy and
pressure (with the divergent vacuum contributions subtracted out).
This, we argue, is the correct version of the sum rule by Karsch,
Kharzeev and Tuchin \cite{Kharzeev:2007wb,Karsch:2007jc}.  The right
hand side can be transformed into Eq.~(\ref{bulkQCD-sr}) by using the
thermodynamic relations $d\epsilon=Tds$ and $dP/d\epsilon=c_s^2$.
Note that Eq.~(\ref{prebulk}) does not coincide with
Refs.~\cite{Kharzeev:2007wb,Karsch:2007jc}, which
had $T\d/\d T$
instead of $3s\d/\d s$.
The issue is the non-commutativity of the $q\to0$ and $\omega\to0$
limits in the bulk channel (see Appendix~\ref{sec:noncommut}).  The
correct expression for the right hand side follows directly from
entropy conservation in hydrodynamics.

In the weak coupling limit of high-temperature gauge theory,
the pressure is given by \cite{Kapusta:1979fh,Kapusta}
\begin{equation}
P=T^4\left(A+ g_s^2 B+{\cal O}(g_s^3)\right)\,,
\end{equation}
where $A,B$ are constant that are unimportant for the following discussion.
Calculating the trace anomaly from $\epsilon=T\frac{d P}{dT}-P$ one finds
\begin{equation}
\epsilon-3P=2 B T^4 g_s \beta (g_s)\sim {\cal O}(g_s^4 T^4)\,,
\end{equation}
which implies \cite{Moore:2008ws}
\begin{equation}
\left( 3s \frac{\d}{\d s}-4\right)(\epsilon-3P)\sim {\cal O}(g_s^6 T^4)\,,
\end{equation}
where $c_s^{-2}=3+\frac{2B}{A} g\, \beta(g_s)$ was used.
On the other hand, the integral over the spectral function
gives a contribution ${\cal O}(g_s^4 T^4)$ at low frequency 
\cite{Moore:2008ws}. In order for our sum rule (\ref{prebulk})
to hold for weakly coupled QCD, the ${\cal O}(g_s^4 T^4)$ contribution must
be canceled to leading-order by the high frequency tail
in the spectral function. There are indications
that this is indeed what is happening in weakly coupled QCD \cite{Simon}.

\section{Conclusion}

In this paper we have written down several sum rules involving the
spectral functions in hot gauge theories.  The sum rules can be
checked in ${\cal N}=4$ SYM theory using gauge/gravity duality.  The
bulk sum rule for QCD was also derived.  We still have some
uncertainty in the shear sum rule in QCD, but hopefully this will be
resolved in the future.

Some conclusions may be drawn from our work.  First, note that
the left-hand side (LHS) of the sum rule (\ref{shearSYM-sr2}) is positive
in strongly coupled ${\cal N}=4$ SYM because $\eta \tau_\pi>\frac{1}{2}\kappa$.
This implies that the spectral function $\delta \rho$ in the shear channel
must be larger than $\eta \omega$ for some frequencies, or otherwise
the integral would not be positive. This feature of the spectral function
is clearly seen on Fig.~\ref{fig:SR}.
This requirement is not satisfied by the 
simplest Lorentzian ansatz for the spectral function,
$\delta \rho(\omega)\sim \frac{\eta \omega}{\omega^2+\Gamma^2}$.
A similar argument can be made for weakly coupled QCD in the large $N_c$ limit, because
there $\tau_\pi\sim 6 \eta/(sT)$ \cite{York:2008rr}
and $\eta/s\sim \alpha_s^{-2} \ln \alpha_s^{-1/2}$ \cite{Arnold:2003zc},
while $\kappa\sim T^2$ (see appendix 
\ref{sec:Ckappa})\footnote{For non-conformal theories, 
corrections to the definition of
$\kappa$ should be suppressed by an additional power of 
$\alpha_s$ \cite{Romatschke:2009im}.}, so one expects
$\eta \tau_\pi>\frac{1}{2}\kappa$.  Therefore, to satisfy both shear
sum rules in QCD, an ansatz more sophisticated than the 
simplest Lorentzian ansatz 
$\delta \rho(\omega)\sim \frac{\eta \omega}{\omega^2+\Gamma^2}$ is needed.

\begin{figure}[t]
\begin{center}
\includegraphics[width=.45\linewidth]{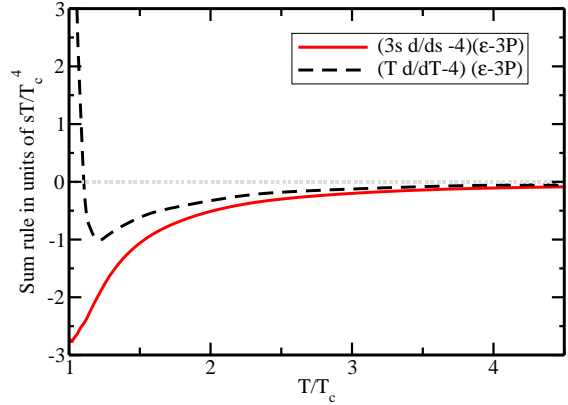}
\end{center}
\caption{The LHS of the bulk sum rule 
[Eqs.~(\ref{bulkQCD-sr}) and (\ref{prebulk})]
(full line), as evaluated from SU(3) lattice data \cite{Boyd:1996bx}.
The result is negative for all temperatures shown, and should not
be associated with the value of the bulk viscosity in SU(3).
For comparison, the result from evaluating Eq.~(\ref{sumrule-q})
(corresponding to the sum rule by Kharzeev and Tuchin \cite{Kharzeev:2007wb})
is shown (dashed lines). The horizontal line is a visual aid to the eye.}
\label{fig:SU3}
\end{figure}

Moreover, the LHS of our sum rule (\ref{bulkQCD-sr}) for the bulk
sector can be evaluated using lattice results for the 
thermodynamics~\cite{Boyd:1996bx}. The result
turns out to be negative for all temperatures above the deconfinement
transition (see Fig.~\ref{fig:SU3})
Fig.~\ref{fig:SU3} demonstrates that the sum rule (\ref{bulkQCD-sr}),
cannot be directly used to extract information about the value of the
bulk viscosity in QCD, unless additional phenomenological assumptions
are made, for example as in
Refs.~\cite{Kharzeev:2007wb,Karsch:2007jc}.  In our language, 
Refs.~\cite{Kharzeev:2007wb,Karsch:2007jc} assume $\rho_{T=0}(\omega)$
to contain a ``non-perturbative'' part 
(associated with the phenomenological gluon condensate), that
---once subtracted--- is offsetting the negative
LHS of the bulk sum rule (\ref{bulkQCD-sr}).



While this work is being completed, we became aware of related work by
Simon Caron-Huot~\cite{Simon}.  We thank L.~Yaffe for discussions,
G.~Moore and S.~Caron-Huot for comments on the manuscript, 
J.~Engels for providing the lattice QCD data used in Fig.~\ref{fig:SU3},
and
S.~Caron-Huot for correcting an error in our earlier treatment of the
bulk sum rule and for giving us a preview of Ref.~\cite{Simon}.  This
work is supported, in part, by U.S.\ DOE grant No.\ DE-FG02-00ER41132.


\appendix

\section{The coefficient $\kappa$}
\label{sec:Ckappa}

Here we calculate directly the Euclidean correlator
\begin{equation}
G(x_1,x_2)=\left<T_{xy}(x_1)\, T_{xy}(x_2)\right>_T
\end{equation}
for a free SU(N) gauge theory at finite temperature $T$. We have
\begin{equation}
G(x_1,x_2)=\left<\left(\partial_x A^\alpha_a-\partial_\alpha A^x_a\right)
\left(\partial_\alpha A^y_a-\partial_y A^\alpha_a\right)
\left(\partial_x A^\beta_b -\partial_\beta A^x_b\right)
\left(\partial_\beta A^y_b-\partial_y A^\beta_b\right)\right>_T\,,
\end{equation}
where each of the building blocks is a correlator of the form
\begin{eqnarray}
C_{i_1 i_2 j_1 j_2 l_1 l_2 m_1 m_2} (x_1,x_2)&=&\left<\partial_{i_1} A^{i_2}_a
\partial_{j_1} A^{j_2}_a \partial_{l_1} A^{l_2}_b
\partial_{m_1} A^{m_2}_b\right>_T\nonumber\\
&&i^4\sumint_{P_1,P_2,P_3,P_4}
e^{i (P_1+P_2)\cdot x_1+i (P_3+P_4)\cdot x_2}  P_1^{i_1}
P_2^{j_1} P_3^{l_1} P_4^{m_1}\nonumber\\
&&\times \left< A^{i_2}_a(P_1) A_{a}^{j_2}(P_2) A_b^{l_2}(P_3) A_b^{m_2}(P_4)
\right>_T\,,
\end{eqnarray}
and $P=({\bf p},p_4)=({\bf p},2 \pi T n)$, 
$\sumint_P=T \sum_{n} \int \frac{d^3 {\bf p}}{(2\pi)^3}$.
We will be interested in the Fourier transform of these correlators,
\begin{eqnarray}
C_{i_1 i_2 j_1 j_2 l_1 l_2 m_1 m_2} ({\bf q},q_4)&=&
\int_0^\beta d\tau \int d^3{\bf x} e^{-i Q\cdot x} 
C_{i_1 i_2 j_1 j_2 l_1 l_2 m_1 m_2} (x,0)\nonumber\\
&=&\sumint_{P_2,P_3,P_4} (Q-P_2)^{i_1} P_2^{j_1} P_3^{l_1} P_4^{m_1}\times
\nonumber\\
&&
\left[\left<A^{i_2}_a(Q-P_2) A_b^{m_2}(P_4)\right>_T\, 
\left<A^{j_2}_a(P_2) A_b^{l_2}(P_3)\right>_T \right.\nonumber\\\
&&\left.
+\left<A^{i_2}_a(Q-P_2) A_b^{l_2}(P_3)\right>_T\, 
\left<A^{j_2}_a(P_2) A_b^{m_2}(P_4)\right>_T\right.\nonumber\\
&&\left.
+\left<A^{i_2}_a(Q-P_2) A_a^{j_2}(P_2)\right>_T\, 
\left<A^{l_2}_b(P_3) A_b^{m_2}(P_4)\right>_T
\right],
\label{longeq}
\end{eqnarray}
where
\begin{equation}
\left<A^{\mu}_a(P_1) A_b^{\nu}(P_2)\right>_T=\frac{1}{T} 
\delta_{n_1+n_2,0} (2\pi)^3
\delta({\bf p_1}+{\bf p_2}) \Delta^{\mu \nu}_{ab}(P_1)\,,
\end{equation}
and $\Delta^{\mu \nu}_{ab}(P_1)$ is the gluon propagator. Note that the
last term in Eq.~(\ref{longeq}) corresponds to a disconnected diagram;
we are only interested in the connected Green's function, so this
term will be dropped in the following.
Using Feynman gauge 
$\Delta^{\mu \nu}_{ab}(P_1)=\delta_{ab} \delta^{\mu \nu} P_1^{-2}$,
Eq.~(\ref{longeq}) simplifies to
\begin{eqnarray}
C_{i_1 i_2 j_1 j_2 l_1 l_2 m_1 m_2} ({\bf q},q_4)&=&(N_c^2-1) \sumint_K
(Q+K)^{-2} K^{-2} (Q+K)^{l_1} K^{m_1}
\times\nonumber\\
&&\left[(Q+K)^{i_1} \delta^{i_2 l_2} K^{j_1} \delta^{j_2 m_2} 
+K^{i_1} \delta^{i_2 m_2} (Q+K)^{j_1} \delta^{j_2 l_2} \right],
\end{eqnarray}
and hence the Fourier transform of the energy momentum correlator becomes
\begin{equation}
\begin{split}
G(Q)=& \,(N_c^2-1) \sumint_K 
(Q+K)^{-2} K^{-2} \times\\
 & \left[4 k_x^2 k_y^2-2 K\cdot (Q+K)(k_x^2+k_y^2)+K^2 k_x^2+(Q+K)^2 k_y^2
+\left(K\cdot(Q+K)\right)^2\right].
\end{split}
\label{GPweak}
\end{equation}

Since we are interested here in the case for vanishing external frequency ($p_4=0$),
the thermal sums are readily evaluated. Dropping the vacuum part 
and using the substitution ${\bf k}\rightarrow {\bf k-q}$ in 
part of the integrand one finds
\begin{eqnarray}
T \sum_n (Q+K)^{-2} &=&\frac{n(k)}{k}\,,\nonumber\\
T \sum_n (Q+K)^{-2} K^{-2} &=&\frac{n(k)}{k}
\left[\frac{1}{|{\bf k}+{\bf q}|^2-k^2}+\frac{1}{|{\bf k}
  -{\bf q}|^2-k^2}\right],\nonumber\\
T \sum_n \frac{K\cdot(Q+K)}{(Q+K)^{2} K^{2}}&=&\frac{n(k)}{k}
\left[\frac{{\bf k}\cdot {\bf q}}{|{\bf k}+{\bf q}|^2-k^2}
-\frac{{\bf k}\cdot {\bf q}}{|{\bf k}-{\bf q}|^2-k^2}\right],\nonumber\\
T \sum_n \frac{\left(K\cdot(Q+K)\right)^2}{(Q+K)^{2} K^{2}}&=&\frac{n(k)}{k}
\left[\frac{({\bf k}\cdot {\bf q})^2}{|{\bf k}+{\bf q}|^2-k^2}
+\frac{({\bf k}\cdot {\bf q})^2}{|{\bf k}-{\bf q}|^2-k^2}\right].
\end{eqnarray}
Expanding the integrand to ${\cal O}({\bf q}^2)$, all the remaining
integrals can be done analytically and one finds
\begin{equation}
\delta G(0,{\bf q})=\frac{N^2-1}{36} T^2 q^2+{\cal O}({\bf q}^4)\,,
\end{equation}
so that
\begin{equation}
\kappa=\frac{N^2-1}{18} T^2\,.
\end{equation}

\section{The coefficient $f_q(0)$ in weakly coupled SU($N$)}
\label{sec:C-pert}

Calculation of $G_R^\infty=\lim_{q_4\rightarrow \infty} G_R(Q)$ starts 
similar to the calculation for $\kappa$ in the previous section, leading
to Eq.~(\ref{GPweak}) for $G_R(Q)$. For ${\bf q}=0$ the sum-integrals become
\begin{eqnarray}
T \sum_n (Q+K)^{-2} &=&\frac{1+2 n(k)}{2 k}\nonumber\,,\\
T \sum_n (Q+K)^{-2} K^{-2} &=&\frac{1+2 n(k)}{2 k}
\left[\frac{1}{q_4^2+2 i k q_4}+\frac{1}{q_4^2-2 i k q_4}\right],\nonumber\\
T \sum_n \frac{K\cdot(Q+K)}{(Q+K)^{2} K^{2}}&=&\frac{1+2 n(k)}{2 k}
\left[\frac{i k q_4}{q_4^2+2 i k q_4}
-\frac{i k q_4}{q_4^2-2 i k q_4}\right],\nonumber\\
T \sum_n \frac{\left(K\cdot(Q+K)\right)^2}{(Q+K)^{2} K^{2}}
  &=&-\frac{1+2 n(k)}{2k}
\left[\frac{k^2 q_4^2}{q_4^2+2 i k q_4}
  +\frac{k^2 q_4^2}{q_4^2-2 i k q_4}\right],
\end{eqnarray}
where we used $n(i q_4)=n(2 \pi i T n)=1$. Evaluation of the remaining
integrals is straightforward and we find
\begin{equation}
\lim_{q_4\rightarrow \infty}G(q_4,0)-G(q_4,0)_{T=0}=f_{q=0}(0)=-2 P,
\end{equation}
where $P=\frac{2}{90} (N_c^2-1) \pi^2 T^4$.
This result can also be obtained by integrating the result for the spectral
function from Ref.~\cite{Meyer:2008gt}.

\section{Non-commutativity of the $\omega\to0$ and $q\to0$ limits of the 
$\<\theta\theta\>$ correlator}
\label{sec:noncommut}

In Sec.~\ref{sec:bulkQCD} we have shown that
\begin{equation}\label{thth-our}
  \lim_{\omega\to0}\lim_{\q\to0} 
  \<\theta\theta\>^{\phantom{1}}_{\!R}(\omega,\q) =
  - \left( 3s\frac\d{\d s}-4\right)(\epsilon-3P).
\end{equation}
Using the same method, we now show that
\begin{equation}\label{ththKT}
  \lim_{\q\to0}\lim_{\omega\to0} 
  \<\theta\theta\>^{\phantom{1}}_{\!R}(\omega,\q) =
  - \left( T\frac\d{\d T}-4\right)(\epsilon-3P).
\end{equation}
This result is consistent with previous results~\cite{Ellis:1998kj}
derived through the Euclidean path integral following the method of
Ref.~\cite{Novikov:1979va}.  It should be expected: Euclidean
correlators are defined with discrete Matsubara frequencies
$\omega_E=2\pi nT$, and the only sensible zero momentum limit in the
Matsubara formalism is to set $\omega_E=0$ first, and then take $\q\to
0$.

We turn on a static metric perturbation,
\begin{equation}
  g_{\mu\nu} = \eta_{\mu\nu} e^{-2\Omega(\bf x)}.
\end{equation}
When $\Omega(\bf x)$ varies smoothly, one can use hydrodynamics to
find out the response.  For static perturbations we expect the
response will be static.  The velocity field is $u^0=e^\Omega$,
$u^i=0$, and the temperature depends on space, $T=T(\bf x)$.
Substituting $T^{\mu\nu}=(\epsilon+P)u^\mu u^\nu+Pg^{\mu\nu}$ into the
equation $\nabla_\mu T^{\mu\nu}=0$, we find
\begin{equation}
   \d_i P - (\epsilon+P)\d_i\Omega =0.
\end{equation}
Using $dP=sdT$ and $\epsilon+P=Ts$, the solution to this equation is
\begin{equation}
  T = T_0 e^{\Omega}.
\end{equation}
The correlator is found from
\begin{equation}
  \<\theta \theta\>(\omega=0,\q\to0) = \frac{\d}{\d\Omega}(\sqrt{-g}T^\mu_\mu)
  = - \left( T\frac\d{\d T}-4\right)(\epsilon-3P).
\end{equation}

One apparent paradox is that if one writes down the bulk sum rule for
any spatial momentum $\q\neq0$, the integral should be equal to
$-\<\theta\theta\>(0,\q)$ which is given by~(\ref{ththKT}) but not
(\ref{thth-our}):
\begin{equation}\label{sumrule-q}
  \left( T\frac\d{\d T}-4\right) (\epsilon-3P)=\frac{2}{\pi}
  \int\limits_0^\infty\!\frac{d\omega}{\omega}\, 
  \delta\rho^{\rm bulk}(\omega,\q)\,,\qquad \q\neq0.
\end{equation}
There is no contradiction, however, as the integral in
Eq.~(\ref{sumrule-q}) is expected to receive a finite contribution
from the region $\omega\sim q$, in particular from the sound-wave peak
at $\omega=c_sq$, as $\<\theta\theta\>$ correlator has a sound-wave
pole.  When $q\to0$, this region shrinks to zero size, but its
contribution remains finite.  The contribution from the sound-wave
peak can be calculated as follows: to leading order in hydrodynamic
fluctuations, $T^{xx}=T^{yy}=T^{zz}=c_s^2 T^{00}$, so 
\begin{equation}
G_R(\omega,q)^{\rm bulk}=(1-3 c_s^2)^2 \<T^{00} T^{00}\>_R\,.
\end{equation}
Defining $\<T^{00} T^{00}\>_R=\frac{q^2}{\omega^2} \chi_L(\omega,q)\simeq \frac{\chi_L(\omega,q)}{c_s^2}$
and using Teaney's result for the spectral density
corresponding to $\chi_L$ \cite{Teaney:2006nc},
\begin{equation}
\frac{\delta \rho_L(\omega,q)}{\omega}=\frac{\epsilon+p}{2}\left[\frac{\Gamma_s\, q^2/2}{
(\omega-c_s q)^2+\left(\Gamma_s\, q^2/2\right)^2}+(\omega\rightarrow -\omega)\right]\,,
\end{equation}
where $\Gamma_s=(\epsilon+P)^{-1}\left(\frac{4}{3}\eta+\zeta\right)$,
the integral over the sound-wave pole at positive frequency gives
$\lim_{q\rightarrow 0}\frac{2}{\pi}
  \int\limits_0^\infty\!\frac{d\omega}{\omega}\, 
  \delta\rho_L(\omega,\q)=\epsilon+P
$. Therefore, the contribution for the integral over $\delta \rho^{\rm bulk}$ is
precisely the difference between the LHS of 
Kharzeev--Tuchin's and our sum rule,
\begin{equation}
  \left(T\frac\d{\d T}- 3s\frac\d{\d s}\right)(\epsilon-3P)
  = \frac{(1-3c_s^2)^2}{c_s^2}(\epsilon+P)\,.
\end{equation}
On the other hand, in the bulk sum
rule~(\ref{prebulk}), we first compute the spectral function
$\rho(\omega)$ at any finite, nonzero $\omega$ by setting $\q=0$ in
$\rho(\omega,\q)$, and then take the spectral integral.  The
sound-wave contribution does not appear in this integral, which means
that our Eq.~(\ref{prebulk}), but not the Kharzeev and Tuchin's version, applies.
Note that at
zero temperature the two limits $\omega\to0$ and $\q\to0$ commute.

\end{document}